\documentclass[a4,times,twocolumn,aps,superscriptaddress]{revtex4-2}
\usepackage{amsmath,amsfonts}
\usepackage{algorithmic}
\usepackage{algorithm}
\usepackage{array}
\usepackage[caption=false,font=normalsize,labelfont=sf,textfont=sf]{subfig}
\usepackage{textcomp}
\usepackage{float}
\usepackage{url}
\usepackage{verbatim}
\usepackage{graphicx}
\usepackage{dcolumn}% Align table columns on decimal point
\usepackage{bm}% bold math
\usepackage[utf8]{inputenc}
\usepackage[T1]{fontenc}
\usepackage{mathptmx}
\usepackage{etoolbox}
\usepackage{xcolor}
\usepackage{soul}
\usepackage{ulem}
\usepackage{cancel}
\begin{document}

%\title[Microwave photoconductance decay system for temperature-dependent charge carrier recombination time measurement]{Microwave photoconductance decay system for temperature-dependent charge carrier recombination time measurement}

%\title{Broadband radiofrequency measurement of time-resolved photoconductivity decay}

%\title{Time-resolved photoconductivity decay measurements with broad-band radiofrequency detection and excitation \textcolor{black}{ photon energy}}
\title{\textcolor{black}{A Versatile System for Photoconductance Decay Measurement Across a Wide Range of Semiconductor Materials}}

\author{András Bojtor}
\affiliation{Department of Physics, Institute of Physics, Budapest University of Technology and Economics, M\H{u}egyetem rkp. 3., H-1111 Budapest, Hungary}
\affiliation{Semilab Co. Ltd., Prielle Kornélia u. 2., 1117 Budapest, Hungary}
\author{Dávid Krisztián} 
\affiliation{Department of Physics, Institute of Physics, Budapest University of Technology and Economics, M\H{u}egyetem rkp. 3., H-1111 Budapest, Hungary}
\affiliation{Semilab Co. Ltd., Prielle Kornélia u. 2., 1117 Budapest, Hungary}

%\author{T. Pinel}% 
%\affiliation{Department of Physics, Institute of Physics, Budapest University of Technology and Economics, M\H{u}egyetem rkp. 3., H-1111 Budapest, Hungary}
%\affiliation{\'{E}cole Nationale Sup\'{e}rieure d'ing\'{e}nieurs de Caen, 6 Bd Mar\'{e}chal Juin, 14000 Caen, France}

\author{Gábor Paráda} 
\affiliation{Semilab Co. Ltd., Prielle Kornélia u. 2., 1117 Budapest, Hungary}

\author{Ferenc Korsós}
\affiliation{Semilab Co. Ltd., Prielle Kornélia u. 2., 1117 Budapest, Hungary}

\author{Sándor Kollarics}
\affiliation{Institute for Solid State Physics and Optics, HUN-REN Wigner Research Centre for Physics, PO. Box 49, H-1525, Hungary}
\affiliation{Department of Physics, Institute of Physics, Budapest University of Technology and Economics, M\H{u}egyetem rkp. 3., H-1111 Budapest, Hungary}
\affiliation{ELKH-BME Condensed Matter Research Group Budapest University of Technology and Economics M\H{u}egyetem rkp. 3, 1111 Budapest, Hungary}

\author{Gábor Csősz}
\affiliation{Department of Physics, Institute of Physics, Budapest University of Technology and Economics, M\H{u}egyetem rkp. 3., H-1111 Budapest, Hungary}

\author{Bence G. M\'{a}rkus}
\affiliation{Stavropoulos Center for Complex Quantum Matter, Department of Physics and Astronomy, University of Notre Dame, Notre Dame, Indiana 46556, USA}
\affiliation{Institute for Solid State Physics and Optics, HUN-REN Wigner Research Centre for Physics, PO. Box 49, H-1525, Hungary}
\affiliation{ELKH-BME Condensed Matter Research Group Budapest University of Technology and Economics M\H{u}egyetem rkp. 3, 1111 Budapest, Hungary}

\author{László Forró}
\affiliation{Stavropoulos Center for Complex Quantum Matter, Department of Physics and Astronomy, University of Notre Dame, Notre Dame, Indiana 46556, USA}

\author{Ferenc Simon}
%\email{simon.ferenc@ttk.bme.hu }
%\homepage{http://goliat.eik.bme.hu/~f.simon/}
\affiliation{Institute for Solid State Physics and Optics, HUN-REN Wigner Research Centre for Physics, PO. Box 49, H-1525, Hungary}
\affiliation{Department of Physics, Institute of Physics, Budapest University of Technology and Economics, M\H{u}egyetem rkp. 3., H-1111 Budapest, Hungary}
\affiliation{ELKH-BME Condensed Matter Research Group Budapest University of Technology and Economics M\H{u}egyetem rkp. 3, 1111 Budapest, Hungary}

\date{\today}% It is always \today, today,

\begin{abstract}
Time-resolved photoconductivity is widely used to characterize non-equilibrium charge-carrier lifetime, impurity content, and solar cell efficiency in a broad range of semiconductors. Most measurements are limited to the detection of reflection of electromagnetic radiation at a single frequency and a single photoexciting light wavelength. We present a time-resolved photoconductivity instrument that enables broadband frequency detection (essentially from DC to $100\ \text{GHz}$), temperature-dependent measurements, and multiple excitation photon energy. The measurement is realized with the help of a coplanar waveguide, which acts as an efficient antenna and whose performance was tested over $10\ \text{MHz} - 10\ \text{GHz}$. The instrument enables the study of surface and bulk charge-recombination specific processes.
\end{abstract}

%\begin{IEEEkeywords}
%Photoconductivity decay, Coplanar Waveguide, Charge-Carrier Lifetime, Temperature-dependence, Wavelength-selective, 
%\end{IEEEkeywords}

\maketitle

\section{Introduction}

The dynamics of photo-induced non-equilibrium charge carriers carry important information about recombination dynamics, impurity content, and recombination mechanism \textcolor{black}{ in semiconductor} materials, which is greatly relevant for both fundamental studies and applications. Contactless, radiofrequency, microwave, and mm-wave reflectometric methods \cite{elsomuPCD, elsomuPCD_2, kunst1, muPCD_korai_DebS} are convenient as these are high-speed (as compared to DC measurements) and non-invasive. 
The characteristic relaxation time of the photoconductivity decay process gives the minority charge carrier lifetime in semiconductors.
%\textcolor{red}{\sout{while the induced change in the reflection provides information about the change in conductivity}}. 
The radiofrequency or microwave photoconductivity decay method is widely used in the semiconductor industry to characterize silicon wafers, monitor electrical parameters at different steps of the manufacturing process and \textcolor{black}{ qualify} the deposited layers and surface passivation \textcolor{black}{\cite{Horanyi-Tutto-Tibor, Sinton1, sinton2}}.

On the fundamental research side, identification of the recombination channels helped to disentangle the so-called Shockley--Read--Hall, Auger-related, or radiative recombination mechanisms in semiconductors\textcolor{black}{\cite{Hall_alapmu}}, or the presence of trap states, which selectively capture one type of charge carrier \textcolor{black}{\cite{Si_recomb_Schroder}}. This carries information about the band structure and shows if excitonic processes play a role\textcolor{black}{\cite{Negyedi_ODMR}}. Concerning applications, the knowledge of the charge-carrier recombination time characterizes sample purity and morphology, which is both a key factor for conventional semiconductor electronics or for energy harvesting and generation \textcolor{black}{\cite{Lami_efficiency_morphology,Lami_efficiency_morphology2, Lami_efficiency_morphology3}}. In addition, the search for new semiconductor materials is a constant quest, which is exemplified by the emergence of novel solar-cell candidates, such as the lead-halide perovskites \cite{Gratzel1, Gratzel2} or by the revolution in power electronics \cite{SiCarbid_PowerDev_review,GaN_PowerDev_review} and topological insulators\cite{Mooretopology, elsotopologikusszigetelo}.

A long charge carrier recombination time and strong photo-response are essential for solar cell application \textcolor{black}{\cite{SolarCell_review}}, the mobility of charge carriers determines the usability of the material as a building block for transistors. The microwave photoconductance decay measurement method presented herein enables measuring the recombination time of semiconductors as a function of temperature with the option of distinguishing surface and bulk effects on the recombination dynamics.

The microwave photoconductance decay measurement is a noncontact, nondestructive method that determines the recombination time of excited charge carriers in semiconductor samples. When optical excitation is applied to a semiconductor, electrons are excited from the valence band to the conduction band leaving holes behind. This change in the concentration of mobile charge carriers in the material leads to a change in conductivity \cite{solyom2}. In the case of microwave photoconductance measurement, the sample is probed with microwave radiation and either the reflected (or transmitted) microwave signal is measured\cite{kunst1, kunst2}, or the sample is placed inside a cavity in which case the change of the quality factor and the shift of the resonance frequency are measured\cite{gyregarami2019ultrafast}. It is generally assumed \cite{muPCD_lauer, Kunst3, kunst1, kunst2} that the connection between the change in conductivity and the change in the reflected microwave signal is linear. By measuring the change in the reflected microwave voltage we can measure the change in the conductivity of the material that is connected to the charge carrier concentration present in the sample. This leads to two possibilities for measurements: pulsed irradiation for the measurement of decay curves and continuous wave (CW) irradiation for the comparison of steady states. If a \textcolor{black}{short pulse} pulsed laser is used the decay of the microwave reflection is connected linearly to the decay of the excited charge carrier concentration leading to the observation of the charge carrier recombination process\cite{Kunst3}. If a CW laser is used and an equilibrium state is reached the charge carrier density in the material is equal to the product of recombination time and generation rate. \cite{BojtorACSPhotonics}  
%\textcolor{black}{\sout{Considering the rate equations it is possible to connect the time dependence of the decay curves and the recombination time as well as the signal level of the steady state measurement and the charge carrier concentration.}} 
Combining the two approaches is also possible using a quasi-steady-state measurement\cite{Sinton1, sinton2, Feriek_QSSPCD_tau_dn}, where the generation rate of charge carriers changes significantly slower than in the case of a decay measurement leading to a local equilibrium at all power levels.

The recombination time of excited charge carriers is dependent on the charge carrier concentration\cite{Feriek_QSSPCD_tau_dn, dn_tau_connection1}. The recombination of excited charge carriers can occur through the following three processes: radiative, Auger, and Shockley--Read--Hall (SRH) recombination\cite{SRH_Auger_rad, Hall_alapmu}. All have different charge carrier density and temperature dependence and are prominent at different charge carrier concentration ranges: at high injection limit the SRH process is not sensitive to $\Delta n$, the radiative recombination process is linear and the Auger process is a quadratic function in $\Delta n$, at low injection limit all three processes are independent of the excited charge carrier density. The charge carrier recombination dynamics can also be influenced by the trapping of charge carriers. In this case, the charge carrier recombination process is limited by the temporal absence of one of the charge carrier types leading to extremely long characteristic times of the recombination. 

The recombination process can be further divided into bulk and surface recombination processes. This is an important distinction since the SRH recombination process is about the recombination of excited charge carrier pairs through energy states in the band gap caused by defects, however, the defects that are present in the bulk and the ones on the surface of a material can be substantially different. In the bulk material, these recombination centers are typically impurity atoms or imperfections in the crystal structure (e.g., vacancies, dislocations, grain boundaries, etc.) while on the sample surface dangling bonds of partially bonded atoms \textcolor{black}{ are responsible for} the recombination of excited charge carriers\cite{Stefan_Reiner_lifetimespectro}. The surface recombination time might be limited by the surface recombination velocity or the characteristic time of the diffusion process\cite{Horanyi-Tutto-Tibor}. %If the following three are satisfied: i) the bulk recombination time is long enough and ii) the diffusion coefficient is high enough for the charge carriers generated in the bulk of the material to reach the surface of the sample, and iii) the recombination velocity at the surface is high, then the recombination is called diffusion controlled surface recombination and the recombination process occures on the surface of the sample, as is the general case for silicon samples\cite{Kunst3, Horanyi-Tutto-Tibor}. \textcolor{red}{VAGY} 
In this second case diffusion limited surface recombination occurs, which is typical for high-purity thin silicon wafers without surface passivation.

In the presented measurement system, we realized a complex tool that can measure the recombination dynamics of the material of interest. With the analysis of the decay curves, one can obtain the recombination time of charge carriers. %and mobility
The measurement system provides the possibility of temperature-dependent measurements and the option to choose the excitation wavelength. The data recorded while changing the temperature may provide information on phase transitions in the material as well as the temperature dependence of charge carrier recombination dynamics. By changing the excitation wavelength the proper excitation of the material can be ensured in regard to the band gap of the material. If the wavelength of the excitation changes while the energy remains sufficient for excitation of charge carrier pairs the charge carrier profile in the material changes drastically. If an excitation wavelength near the band gap is applied the charge carriers are generated throughout the whole material. In contrast, with increasing excitation energy the absorption is present more and more at the surface of the material. With this, we are able to measure the recombination time of charge carriers generated at the surface and in the bulk material.

\section{Method of detection}

\subsection{Microwave photoconductance measurement setup}

%\begin{figure}[!t]
\begin{figure}[htb]
    \centering
    \includegraphics[width=0.48\textwidth]{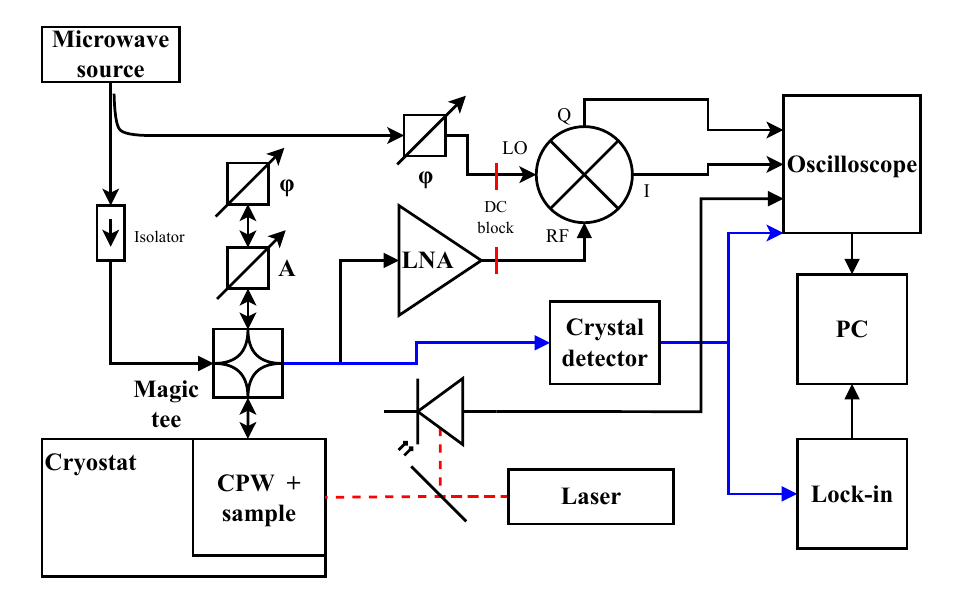}
    \caption{Schematics of the experimental setup showing both the time-resolved measurement option with the IQ mixer and oscilloscope and the option using a lock-in amplifier for the CW measurement (denoted with blue lines). A photodiode is used to trigger the time-resolved photoconductance traces.}
    \label{setup}
\end{figure}
%Here we present a device built for the measurement of microwave photoconductance in multiple possible configurations. 
\noindent
The block diagram of the measurement setup is presented in Figure \ref{setup}. The system is capable of conducting measurements of time-resolved microwave-detected photoconductance decay as well as the measurement of steady states realized during continuous wave excitation. A further possibility to extend the technique toward scanning capabilities is to add a linear stepper motor, thus realizing a one-dimensional mapping of the photoconductance signal. The excitation wavelength (at present $532$ or $1064\ \text{nm}$) can also be selected with ease. As the sample is placed inside a closed-cycle cryostat, it is possible to change the temperature from $10\ \text{K}$ up to room temperature, while there are cryostats available that could go down to $1.5\ \text{K}$ and up to $800\ \text{K}$ with similar geometry. It is also possible to use multiple probing frequencies, below we demonstrate the operation for microwave radiation in the X-band (around $10\ \text{GHz}$) and a radiofrequency one at $50\ \text{MHz}$.

For the microwave detection option, the microwave source (MKU LO 8-13 PLL, Kühne GmbH) provides the probing radiation at a fixed frequency. The signal is split with a directional coupler (R433721, Microonde) between the local oscillator (LO) input of the IQ mixer (through a phase shifter) and towards the sample. The reflectometry measurement is achieved using a Magic Tee duplexer whose reference arm is closed with an attenuator and another phase shifter. The latter setup allows to null (or cancel) any unwanted reflection from the direction of the sample. 
The phase and amplitude of the signal reflected from the reference arm are set in a manner that cancels out the reflection caused by the sample (or any microwave parts) without optical excitation to ensure that the unwanted saturation of the IQ mixer is prevented. Since this allows for a zero-indication measurement, a low-noise preamplifier (NF=2 dB, JaniLab Inc.) can be placed before the mixer, which further increases the signal-to-noise ratio. A disadvantage of this nulling setup is that some readjustment is required during operation.

The Magic Tee is known to reflect half of the incoming power back to the source, which is dissipated using a microwave isolator in front of the Tee. The microwaves are directed with semirigid copper cables toward the sample, which is on a coplanar waveguide on close thermal contact with the cold finger of the cryostat (M-22, CTI-CRYOGENICS) in vacuum. 

The signal is connected to the RF input of an IQ mixer (IQ0618LXP, Marki Microwaves) that provides the detection of the signal split to in-phase and out-of-phase components. The outputs of the IQ mixer are connected to a high-speed oscilloscope (Tektronix MDO3024) thus realizing time-resolved measurements. Both the RF and LO inputs of the mixer are galvanically isolated from the rest of the circuit to prevent ground loops using DC-blocks, which are formed by two pairs of facing SMA-WR90 connectors, which are separated by insulating sheets and plastic screws.

The system can be set up with a crystal detector (8472B, Hewlett Packard) instead of the IQ mixer. This unit rectifies the probing signal and thus enables the utilization of frequencies in the $10\ \text{MHz}\ - \ 18\ \text{GHz}$ range but the phase-sensitive detection is lost. We used a signal generator (SDG 1050, Siglent Inc.) to provide a signal source of $50\ \text{MHz}$ and replaced the microwave Magic Tee with a corresponding RF version (range 1-200 MHz, Anzac HH107). 

The light source is a pulsed laser (NL201-2.5k-SH-mot, Ekspla) which allows for a quick selection between the fundamental $1064\ \text{nm}$ or the frequency doubled $532\ \text{nm}$ emission using an external frequency doubling unit. The laser delivers maximal pulse energies of $0.9\ \text{mJ}$ at $1064\ \text{nm}$ and $0.3\ \text{mJ}$ at $532\ \text{nm}$ emission with tunable repetition rate between $0$ and $2500\ \text{Hz}$. 
A photodiode, placed after a beam sampler, provides the trigger signal for the oscilloscope to measure the photoconductance traces.

In addition, the setup can operate in a continuous wave (CW) option, using separate CW $532\ \text{nm}$ (MGL-III-532-$200\ \text{mW}$) and $1064\ \text{nm}$ (Coherent Adlas DPX301LLOEM) laser with a maximal output of $500\ \text{mW}$ whose intensity is modulated with an optical chopper (SR542, Stanford Research Systems). All these lasers can be combined using efficient optical multiplexing techniques using $50$\% beam splitters. Given that the optical chopping frequency (typically 100 Hz-4 kHz) is much slower than any recombination dynamics in the sample, this measurement option represents the study of the quasi-steady-state. The reflected signal either from the mixer or from the detector can be measured using a lock-in amplifier (SR830, Stanford Research Systems). 

The sample is placed on a coplanar waveguide, CPW\cite{pozar}, that is placed inside a cryostat, which allows for temperature-dependent measurements between $10\ \text{K}$ and room temperature. The back side of the CPW is connected to the ground and is in good thermal contact with the cold finger of the cryostat. The CPW in our case is essentially a broad-band radiofrequency antenna that operates from DC until the cut-off frequency of our system (about 20 GHz in our case), which is usually characterized by the dielectric material of the microwave components.

\subsection{The method of detection}

%\begin{itemize}
%    \item sorrend legyen Zs -> gamma -> delta gamma
%    \item legyen egy kis bevezető a kábel RLCG leírásáról és ennek perturbálásáról
%\end{itemize}

\begin{figure}[!t]
    \centering
    \includegraphics[width=0.99\linewidth]{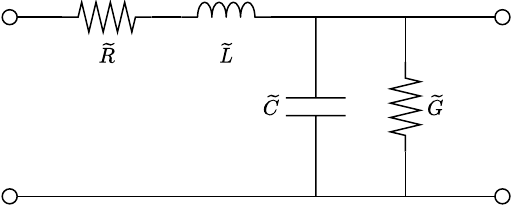}
    \caption{Equivalent circuit of the coplanar waveguide per a unit length.}
    \label{CPWcirc}
\end{figure}
\noindent
As mentioned above, the studied samples are placed on the top of the coplanar waveguide, effectively interacting with its electromagnetic field. The equivalent circuit model of the CPW consists of lumped inductors and capacitors besides the series resistance and shunt conductance, which is shown in Figure \ref{CPWcirc}. The respective parameters (values normalized to unit length) are $\widetilde{L}$, $\widetilde{C}$, $\widetilde{R}$, and $\widetilde{G}$. Parameters $\widetilde{R}$, and $\widetilde{G}$ represent loss and are neglected in ideal cases. In the general case, the wave impedance of the CPW reads\cite{pozar}:
\begin{equation}
    Z_{0}=\sqrt{\frac{\widetilde{R}+\mathrm{i}\omega \widetilde{L}}{\widetilde{G}+\mathrm{i}\omega \widetilde{C}}},
    \label{}
\end{equation}
where $\omega$ is the angular frequency of the radiation. Our CPW is designed for the conventional $Z_0=50\ \Omega$ wave impedance. 

The CPW is perturbed by the presence of the sample, which leads to the measured reflection. The sample is characterized by the surface impedance, or $Z_{\text{s}}$\cite{pozar, chen2004microwave}:
\begin{equation} \label{Zs}
    Z_{\text{s}}=\sqrt{\frac{\mathrm{i}\omega\mu}{\sigma+\mathrm{i}\omega\epsilon}},
\end{equation}
where $\mu=\mu_0 \mu_{\text{r}}$ ($\mu_0$ is the vacuum permeability), $\sigma$ is the sample conductivity, and $\epsilon=\epsilon_0 \epsilon_{\text{r}}$ ($\epsilon_0$ is the dielectric constant). The 'r' subscripts denote the respective material-dependent relative permeability and permittivity.

If a sample is placed onto the CPW, the perturbed surface impedance takes the form $Z_{\text{perturbed}}=\eta Z_\text{s}$ (Refs. \cite{chen2004microwave, BoothEtal_surfaceIMP_waveIMP_connection}), where the $\eta$ \emph{filling factor} is a geometric factor dependent on the sample shape, size, and its coverage of the CPW surface. There are two limiting cases of $Z_\text{s}$: these are the low conductivity ($\sigma\ll\omega\epsilon$, which occurs for a usual dielectric) and the high conductivity limit ($\sigma\gg\omega\epsilon$, also known as the quasi-stationary limit). In both cases, one part of the denominator disappears leading to $Z_\text{s}=\sqrt{\frac{\mathrm{i}\omega\mu}{\sigma}}=\frac{1+\mathrm{i}}{2}\mu\omega\delta$ in the high conductivity range ($\delta=\sqrt{\frac{2}{\mu \omega \sigma}}$ is the penetration depth), and $Z_\text{s}=\sqrt{\frac{\mathrm{i}\omega\mu}{\mathrm{i}\omega\epsilon}}=\sqrt{\frac{\mu}{\epsilon}}$ in the low conductivity range. The latter result can be rewritten with the help of the wave impedance of the vacuum, $Z_{\text{v}}=\sqrt{\frac{\mu_0}{\epsilon_0}}\approx 376.7\,\Omega$ as $Z_\text{s}=\frac{Z_{\text{v}}}{n}$, where $n=\sqrt{\mu_{\text{r}}\epsilon_{\text{r}}}$ is the refractive index.

The connection between the exciting ($U_{\text{exciting}}$) and reflected ($U_{\text{reflected}}$) microwave voltage is given by the reflection coefficient, $\Gamma$. It is connected to the perturbation of the wave impedance ($Z$) caused by the sample placed on the CPW. We note that when the sample does not properly terminate the waveguide the presence of transmitted power (or $S_{21}$) modifies this formula\cite{pozar}. The value of the reflection coefficient is\cite{pozar}:
\begin{equation}\label{}
    \Gamma=\frac{U_{\text{reflected}}}{U_{\text{exciting}}}=\frac{Z_{\text{perturbed}}-Z_0}{Z_{\text{perturbed}}+Z_0}.
\end{equation}
%In this study we assume that there is no transmission of the probing radiation and thus the reflection amplitude, $\text{S}_{11}$, is equal to the magnitude of the reflection coefficient.
Although we keep the discussion as general as possible, we discuss below that our experiment, on a few mm-sized silicon wafers, is well described by $Z_{\text{perturbed}}=Z_{\text{s}}$ (i.e., $\eta=1$). Measurements on very small-sized samples \textcolor{black}{below the size of the gap between the signal and ground strips} may show deviation from this, and a filling factor much smaller than $1$ can be included.

\begin{figure}[!t]
    \centering
    \includegraphics[width=0.99\linewidth]{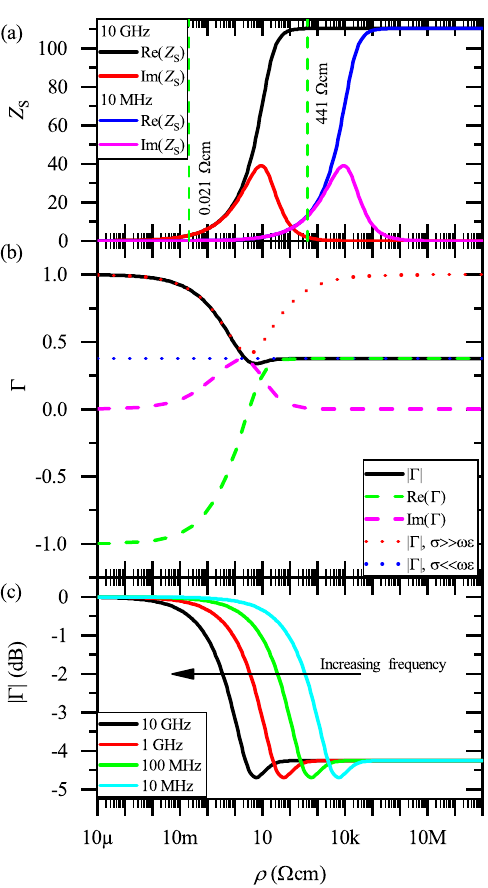}
    \caption{Resistivity dependence of the surface impedance (a) and the reflection coefficient at different probing frequencies. The latter data are shown for both linear and logarithmic scales (b and c).}
    \label{simulation}
\end{figure}

In Figure \ref{simulation}., the reflection coefficient is shown as a function of the sample resistivity at different probing frequencies. Figure \ref{simulation}a. shows the complex surface impedance for two detection frequencies. Figure \ref{simulation}b. depicts the complex reflection coefficient, which is shown for the low conductivity, high conductivity and general cases. Finally, Fig. \ref{simulation}c. gives the probing frequency dependence of the reflection coefficient on a logarithmic scale. In the calulcations, we considered $\epsilon_{\text{r}}=11.68$ as the dielectric constant of silicon\cite{Si_epsilonR_1}.

The calculation shows that a 4 dB change in the reflection coefficient is expected when the sample changes its resistivity between the low- and high-conductivity ranges upon light excitations. A careful examination of the detected signal revealed that we indeed observe similar changes in the reflection which confirms that the above description with $\eta\approx 1$ is valid for our geometry. This consideration included the source power, insertion losses due to the directional coupler (3 dB), the cables (about 5 dB), the magic Tee (6 dB), and eventually due to the conversion loss in the mixer (7.5 dB according to its specifications).

%{\color{red}vhova majd be kene irni hogy ezek tipikus ellenallasertekekt sziliciumban, ill hogy ezek sziliciumban hogyan viszonyulnak a delta n-hez!!!} \textcolor{black}{beírtam 1 bekezdéssel lejjebb. tiszta Si: 5e22 atom/cm3. intrinsic carrier conc: 1e10. agyondópolt: 1e21/cm3, ntype 0.0001ohmcm, ptype 0.0002. kb nincs is benne: 1e12/cm3, ntype 4407.143 ohmcm, ptype 13281.1424ohmcm}

As shown in Figure \ref{simulation}c., the regime of linear response between the reflection coefficient and the resistivity depends on the probing frequency. Clearly, a tuneable probing frequency is required to efficiently obtain the largest possible dynamic range in the reflectometry measurement for a broad range of sample resistivity. In fact, the technique presented herein offers this possibility. Our phase-sensitive detection method allows to monitor the phase changes in the reflected microwave voltage, too. This leads to the possibility of the measurement of the complex material parameters in the case of a proper calibration of the phase for the reflected microwave signal.

In intrinsic (or undoped) silicon, the charge carrier concentration is around $10^{10}\ \frac{1}{\text{cm}^3}$. Typical doping levels range from $10^{13}\ \frac{1}{\text{cm}^3}$ to $10^{18}\ \frac{1}{\text{cm}^3}$, doping levels above $10^{18}\ \frac{1}{\text{cm}^3}$ leads to degenerate doping. The resistivity of silicon at a doping level of $10^{13}\ \frac{1}{\text{cm}^3}$ is $\rho\approx441\ \Omega\text{cm}$ in $n$-type and $\rho\approx1330\ \Omega\text{cm}$ in $p$-type samples at room temperature. With a doping level of $10^{18}\ \frac{1}{\text{cm}^3}$ the resistivity is $\rho\approx0.021\ \Omega\text{cm}$ in $n$-type and $\rho\approx0.0366\ \Omega\text{cm}$ in $p$-type samples. These values support that the resistivity-dependent curves shown in Figure \ref{simulation}. are in the range of typical resistivity values and that a proper probing frequency can be selected to ensure the monotonicity of the reflection coefficient during the measurement process.

The probing frequency also determines the penetration depth of the radiofrequency radiation into the material as 
$\delta=\sqrt{\frac{2}{\mu \omega \sigma}}$. This leads to the conclusion that the lower the frequency of the probing electromagnetic wave, the higher the penetration depth is. This results in a penetration depth of $\delta=1.59\ \text{mm}$ for a probing frequency of $10\ \text{GHz}$ and $\delta=50.34\ \text{mm}$ for $10\ \text{MHz}$ in the case of a silicon sample with $10\ \Omega\text{cm}$ resistivity. The penetration depth decreases with decreasing sample resistivity, thus changing the selective probing effect. This shows that the penetration depth is approximately two times the sample thickness ($0.75 \text{mm}$) in the dark for a probing frequency of $10\ \text{GHz}$ and decreases as a result of excitation.

The reflected voltage and sample conductivity are generally assumed to be in a linear connection\cite{muPCD_lauer, kunst1}. Our calculation above shows that this is valid for a broad range of charge carriers, either due to doping or due to light-induced excitation, provided the appropriate probing frequency is chosen. Furthermore, we employ the connection between the conductivity and charge carrier density given by the Drude model\cite{solyom2}:
\begin{equation}\label{}
    \sigma=\frac{n_{\text{e}} e^2\tau}{m_e^{\ast}}=n_{\text{e}} e \mu_\text{e},
\end{equation}
where $n_{\text{e}}$ is the electron density, $e$ is the elementary charge, $\tau$ is the momentum-scattering time,  $m_e^{\ast}$ is the effective electron mass or band mass, and $\mu_\text{e}$ is the electron mobility.

When the linearization approach is valid \cite{kunst1, kunst2, Kunst3, muPCD_lauer}, we calculate the Taylor series of the surface impedance and reflection coefficient for small conductivity changes as:
\begin{equation}\label{}
    Z_\text{s}\left(\sigma +\Delta\sigma\right)\approx Z_\text{s}\left(\sigma\right)\cdot\left(1-\frac{\Delta\sigma}{2\left(\text{i}\omega\epsilon+\sigma\right)}\right),
\end{equation}

\begin{equation}\label{}
	\Gamma\left(\sigma+\Delta \sigma\right) \approx \Gamma_0 \left(1-\frac{Z_0\cdot Z_\text{s}\left(\sigma\right)}{Z_\text{s}\left(\sigma\right)^2-Z_0^2}\cdot\frac{\Delta \sigma}{\text{i}\omega\epsilon+\sigma}\right).
\end{equation}

Using these, we obtain the connection between the change in the reflected microwave voltage and the change in charge carrier density as\cite{kunst1, kunst2}:
\begin{equation}\label{}
	\Delta \Gamma \propto\Gamma_0\cdot \Delta \sigma\propto \Delta n.
\end{equation}

%This connection can be easily shown by analyzing the Taylor series of the reflection coefficient. If the change in conductivity is small the connection is as follows:

\subsection{Samples}

\noindent
To demonstrate the properties of the measurement setup, we chose silicon samples. We used a \textcolor{black}{moderate resistivity} silicon sample \textcolor{black}{typically used for IC fabrication} to demonstrate the sensitivity of the coplanar waveguide-based approach. The sample used for the comparison of probing wavelength and phase, waveguide type, and excitation wavelength is a larger $n$-type, $\sim 18.5\ \Omega \text{cm}$ silicon sample of approximately $10\ \text{mm }\times10\ \text{mm}$. We removed the native oxide from the sample surface with hydrofluoric acid. We chose silicon samples because they are widely investigated leading to a straightforward comparison of our results with the literature values.

\section{Characterization and performance of the instrument}
\noindent
To characterize the measurement setup, we studied the sensitivity profile of the coplanar waveguide, the phase sensitivity of the detection, and the effects of the probing radiofrequency and light excitation wavelengths. All measurements were done on silicon samples cut out from silicon wafers. We also demonstrate that the instrument allows to study the temperature dependence of the charge-carrier lifetime.

\subsection{Spatially resolved $\mu\text{PCD}$ studies}
\noindent
The study of spatially resolved $\mu\text{PCD}$ arises from the fact that light-induced non-equilibrium charge carriers distribute unevenly in a sample due to charge diffusion. Another effect is that charge recombination is often more rapid near the surface than in the bulk depending on the surface quality\cite{Si_surface_passivation}. A well-known example, when charge-diffusion effects could be disentangled from carrier drift is the Haynes--Shockley experiment\cite{HaynesShockleyExp, HaynesShockleyDiffusion} by means of a spatially resolved $\mu\text{PCD}$ study.

\begin{figure}[!t]
    \centering
    \includegraphics[width=0.99\linewidth]{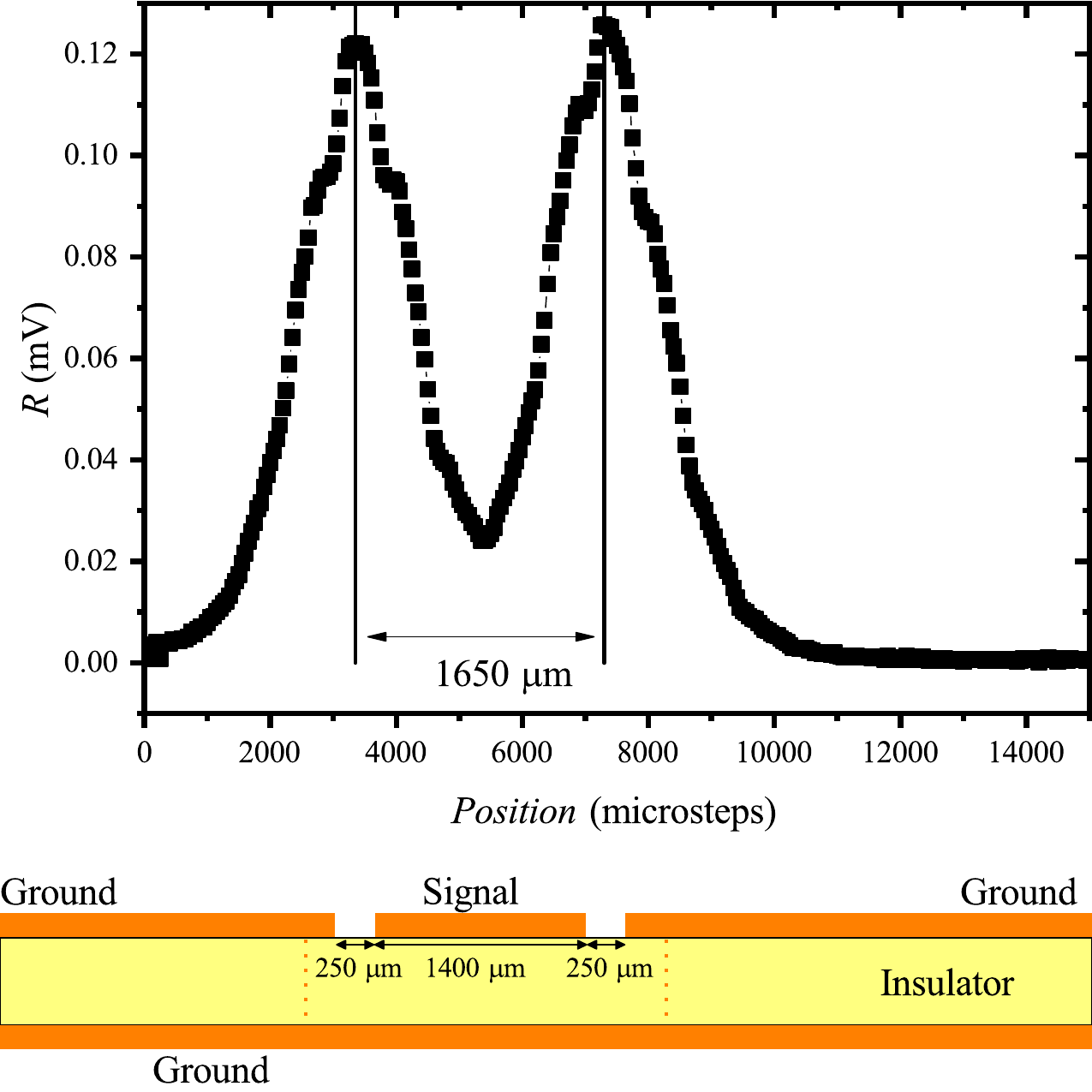}
    \caption{Spatially resolved microwave photoconductance signal measured using a stepper motor and lock-in amplifier. The amplitude of the detected signal, $R$, is shown. The bottom image shows the layout of the CPW with respect to the actual measurement on a 1:1 scale.}
    \label{leptetomotor}
\end{figure}

The exceptional spatial sensitivity of our instrument stems from the geometry of the CPW: the microwave field is localized to the gaps between the center conductor and the grounded side conductors. 
Figure \ref{leptetomotor}. shows the result of a measurement where the exciting laser spot was scanned on the surface of the sample with the help of a stepper motor (TMCM-1070, Trinamic). During this measurement we used a $4\ \text{mW}$, $\lambda=670\ \text{nm}$ CW laser (Thorlabs CPS186) and applied a modulation to the driving current, in other cases, chopping is also an option. The laser beam was focused using a lens down to a diffraction-limited focus size with a diameter of about $d=12\, \mu\text{m}$. This is obtained from the Abbe formula of $d=\lambda/2 \text{N.A.}$, N.A. being the numerical aperture, which was estimated from the exciting beam waist diameter and the lens-sample separation. The signal was recorded with the help of a lock-in amplifier placed after a crystal detector. We use a homogeneous silicon sample with high resistance and low recombination time ($\approx 8.5\,\mu\text{s}$), which gives 
a charge diffusion length of $\approx175\ \mu\text{m}$ for electrons and $\approx101\ \mu\text{m}$ for holes (the difference is due to the differing mobility).

Given that both the laser spot size and the charge diffusion length are smaller than the characteristic length scale of the CPW (about $250\,\mu\text{m}$), we expect that the observed photoconductance signal measures the sensitivity profile of the CPW. Figure \ref{leptetomotor}. shows that the maximum signal is observed at the points above the gap between the central and side conductors. Thus, it is ideal to place the sample to this location for maximum sensitivity. The separation between the signal maxima is $1650\ \mu\text{m}$ which is explained by the width of the central conductor ($1400\ \mu\text{m}$) and the CPW gaps ($250\ \mu\text{m}$). The signal decays rapidly toward the side of the CPW; while this is in good agreement with our expectation, it shows an exceptional spatial resolution. Radiofrequency antenna configurations typically deliver a spatial resolution that is comparable to the wavelength. However, the spatial resolution for the CPW is determined by the gap size, therefore our design is capable of delivering far superior resolution than conventional approaches. This is particularly true for lower frequencies, as the wavelength of the probing radiation, e.g., at 10 MHz is as large as 30 meters. The spatial resolution of the measurement is further limited by the diffusion length of excited charge carriers, $\sim100\ \mu\text{m}$ in this measurement. 

One could also place a sample parallel to the gaps of the CPW above one gap and scan the laser spot above the gap. The final limitation on spatial resolution during such a measurement is the lowest value between the diffusion length, step size of the stepper motor, and the spot size of the laser.

\subsection{The phase sensitive detection of $\mu\text{PCD}$}
\noindent
The phase of the reflected radiofrequency signal from a sample contains information about the complex material parameters and thus a phase-sensitive detection is superior to the detection of reflected power only. If a sample with sufficient thickness is measured, the phase of the reflected microwave radiation also contains the information of the depth that the reflection occurs at, since traveling through a material causes a phase shift in the measurement.

\begin{figure}[!t]
    \centering
    \includegraphics[width=0.99\linewidth]{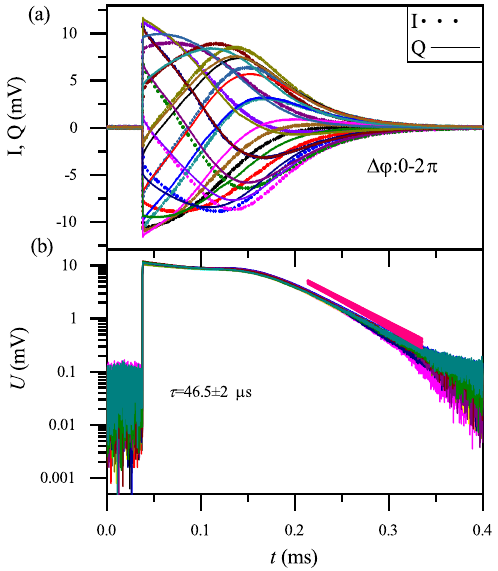}
    \caption{a) A series of measurements with different phase differences between the LO and RF arm of the IQ mixer demonstrate that the signal magnitude is insensitive to the setting. b) The time decay of the magnitude is shown with an offset applied to the fitted curves in pink. The results show that there is no substantial difference between subsequent measurements with different phases, leading to correct measurements despite the phase change.}
    %b) The time decay of the magnitude is , the magnitude of the measured photoconductance signal is shown on subfigure (b). 
    \label{LOphase}
\end{figure}

Figure \ref{LOphase}. shows the photoconductance decay measurement with multiple phase differences between the LO and RF input of the IQ mixer. We created the magnitude of the detected signal by taking the square root of the square of the product of the two channels, $U=\sqrt{\text{I}^2+\text{Q}^2}$. The I and Q signals represent the complex reflection coefficient. The results show that no matter how different the I and Q channels are, the result remains the same. This proves that the IQ mixer works according to our expectations, the phase difference between the LO and RF arm does not cause any problem with the evaluation process. This also shows that the result is physically relevant despite the I or Q channels showing sign changes and non-monotonic behavior. Measurement of the signal magnitude disregards the information carried in the phase.

\subsection{Broadband $\mu\text{PCD}$ measurement using the coplanar waveguide}
\noindent
Our measurement setup is capable of measurements in a wide range of probing frequencies. We can set values essentially in the DC-$18\ \text{GHz}$ range, while in practice we only tested operations from $10\ \text{MHz}$. However, we do not have a phase-sensitive mixer for the low-frequency range, we thus demonstrated the operation with a crystal detector and a signal generator. The measurement conducted with different probing frequencies lead to a difference in the surface impedance as shown in Equation \eqref{Zs}. This leads to the difference in resistivity-dependent reflection shown in Figure \ref{simulation}. The simulation shows the reflection as a function of resistivity for multiple different probing frequency values. For an ideal measurement, one would set the frequency in a manner that the connection between reflection and resistivity is linear or can be linearized on the whole measurement range. The frequency of the probing radiofrequency radiation also determines the penetration depth of the radiation into the sample.

\begin{figure}[!t]
    \centering
    \includegraphics[width=0.99\linewidth]{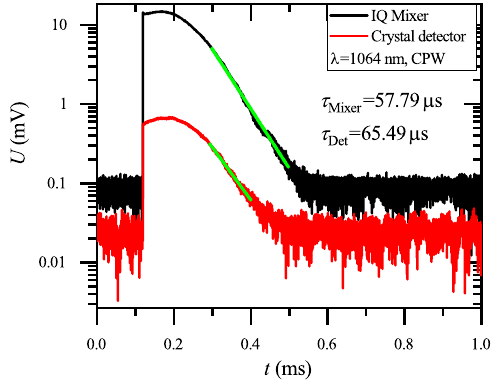}
    \caption{Microwave photoconductance decay curves recorded using $50\ \text{MHz}$ probing frequency with a crystal detector and $10\ \text{GHz}$ probing frequency with an IQ mixer. The similar recombination times show that we are in the linear regime of the measurement for both probing frequencies.}
    \label{broadband-measurement}
\end{figure}

Figure \ref{broadband-measurement}. shows the result from the two measurement methods, with a crystal detector and an IQ mixer. For the latter, the signal magnitude is shown. The results obtained during our measurement show minimal difference in the time traces of the measurement and the charge-carrier recombination time. This also means that the conductivity range reached during the measurement is such that it falls in the linear regime of the reflected signal vs charge carrier density relation for both frequencies.

%\subsection{The different probing of coplanar and rectangular waveguides}

%\begin{figure}[!t]
%    \centering
%    \includegraphics[width=0.99\linewidth]{Graph_cpw-rect.png}
%    \caption{Coplanar VS rectangular waveguide}
%    \label{fig:enter-label}
%\end{figure}

%\begin{itemize}
%    \item egy kép kellene még a CPW és rectangular waveguide bemutatásáról
%\end{itemize}

\subsection{Laser wavelength selective $\mu$PCD studies}
\noindent
The wavelength of excitation is an important parameter during the investigation of recombination processes. The absorption coefficient and thus the absorption depth depend greatly on the wavelength of illumination. This leads to a wavelength-dependent location of the photoexcited charge carriers. If an illumination with energy close to the band gap of the material is used, the excitation is more homogeneous throughout the material. However, for higher photon energies the excitation occurs closer to the surface of the material. This effect also depends on the thickness of the material, since for thin layers high energy photons can also cause a homogeneous excitation while for thick enough samples not even excitation near the band gap can excite the material in its whole cross section. The surface quality can also influence the emerging charge carrier distribution, prior to the presented measurements the sample under investigation was submerged in hydrofluoric acid to remove native oxide.

Herein, we compare the results of $532\ \text{nm}$ and $1064\ \text{nm}$ excitation. The band gap of silicon corresponds to an optical wavelength of $1102.9\ \text{nm}$ \cite{Si_bandgap} at room temperature. The absorption lengths are $1.27\ \mu\text{m}$ and $900\ \mu\text{m}$ for the $532\ \text{nm}$ and $1064\ \text{nm}$ excitations, respectively\cite{Si_absorptioncoeff}. Since the sample under investigation is $750\ \mu\text{m}$ thick, the excitation mostly happens near the sample surface for the $532\ \text{nm}$ excitation, whereas charge carriers are excited in the whole sample for the $1064\ \text{nm}$ excitation. \textcolor{black}{ The absorption of light can influence the recombination dynamics of charge carriers in the volume of excitation.}

%This leads to a difference in the photoexcited state between the two applied excitation wavelengths.

\begin{figure}[!t]
    \centering
    \includegraphics[width=0.99\linewidth]{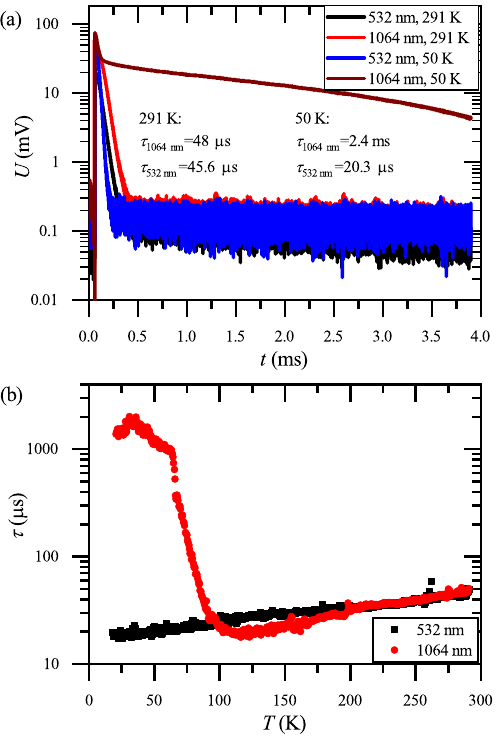}
    \caption{Wavelength-dependent recombination dynamics in the case of near-band gap and high-energy excitation. The temperature-dependent recombination times indicate a difference between bulk and surface effects. Subfigure (a) shows the photoconductance decay curve in the case of room temperature and $50\ \text{K}$ measurement. Subfigure (b) shows the change in recombination time throughout the temperature range for both excitation wavelengths.}
    \label{wavelengthdep}
\end{figure}

With the help of a cryostat, that is capable of changing the temperature in the $20-300\ \text{K}$ range, we conducted temperature-dependent photoconductance decay measurements with both the $532\ \text{nm}$ and the $1064\ \text{nm}$ excitations. As mentioned above, the CPW-based radiofrequency circuit is ideally suited for cryogenic measurements as the CPW sits on the cold finger of a closed-circuit cryostat with good heat anchoring. We analyzed the individual photoconductance decay curves and obtained the characteristic time of the recombination process at each temperature in the range. The result is shown in Figure \ref{wavelengthdep}. 

\textcolor{black}{The data taken with the $532\ \text{nm}$ laser shows a temperature dependence which is characteristic for a diffusion limited recombination. In thin Si wafers, the charge carrier lifetime is sufficiently long that the charge carriers can diffuse to the surface. Then, recombination occurs on the surface and the lifetime is essentially limited by the time it takes for the charges to diffuse to the surface. At low temperatures, the increasing mobility enhances the diffusion, which in turn lowers the charge-carrier lifetime.}
%The data taken with the $532\ \text{nm}$ laser shows a temperature dependence which is characteristic for the so-called Shockley--Read--Hall recombination, or SRH process \textcolor{black}{\cite{SRH_tempdep_NSi, SRH_Tempdep_PSi, Stefan_Reiner_lifetimespectro}}. The SRH recombination or impurity-assisted recombination occurs on deep-level centers which capture both a hole and an electron thus assisting the recombination. The SRH process related charge-carrier recombination time usually lengthens \cite{SRH_tempdep_NSi, SRH_Tempdep_PSi, Stefan_Reiner_lifetimespectro} with increasing temperature as the thermally activated occupation of the impurity levels hinder the SRH process. This fully explains the observed temperature-dependent $\tau_{\text{c}}$ at the $532\ \text{nm}$ excitation: therein the SRH centers near the surface are responsible for the observed temperature dependence. 

However, in general, the bulk of a semiconductor can be very different from the surface. This is the reason why the literature distinguishes the so-called bulk recombination time\cite{Stefan_Reiner_lifetimespectro} which in general can be very different from the overally observed recombination time.

Figure \ref{wavelengthdep}. indeed shows characteristically different behavior when excited with the $1064\ \text{nm}$: below $100\ \text{K}$, $\tau_{\text{c}}$ lengthens enormously to a maximum value of a few milliseconds. Our focus herein is on describing the capabilities of our instrument development thus a detailed analysis of this effect is beyond our scope.
\textcolor{black}{ As a possible explanation, we} propose the difference in behavior to be attributed to the difference in absorption coefficient for the two excitation wavelengths used in this study. In the case of the near-bandgap excitation, the charge carrier generation happens throughout the whole material. In the case of the $532\ \text{nm}$ excitation, the absorption occurs in a significantly \textcolor{black}{ shallower depth} with photons possessing higher energy. The optical excitation may influence the charge states of recombination centers or donor/acceptor atoms in the volume of absorption. This leads to vastly different recombination dynamics throughout the bulk of the material depending on the laser wavelength. \textcolor{black}{In the literature we haven't found the observation or description of similar light energy sensitive behavior.} %Dávid javasata alapján utolsó mondat átírva
%Since optical excitation may influence the recombination dynamics in the volume of absorption the recombination dynamics could be vastly different in the thin region near the sample surface and throughout the bulk of the material.
%\sout{We nevertheless propose that the capture of either an electron or a hole in a level near the respective conduction or valence band, a process known as charge trapping, is responsible for this effect. The charge trapping hinders the charge-carrier recombination through two effects: first, a captured charge is most probably localized thus hindering the diffusion of its charge carrier pairs with the opposite charge to the surface, where the recombination would otherwise happen. Second, the trapped charges have to escape from the trap states in order that they can participate in the recombination. The combination of these effects is known to significantly slow down recombination in silicon at low temperatures}\cite{SiTauTempDep}.%a cite is törlendő lenne, csak hibát dob ha azt áthuzatom.

Charge-carrier concentration of doped semiconductors is temperature dependent with \textcolor{black}{partial ionization or so called freeze out} region at low temperatures\cite{solyom2, Stefan_Reiner_lifetimespectro, SiTauTempDep}, where the dopants are not fully ionized. In the case of $n$-type silicon, this temperature region is reached around $100\ \text{K}$\cite{Si_CarrierFreezeout}. In the freeze-out region, the ionization of dopants is partial at most, which leads to the possible ionization of the dopant atoms with optical excitation that leaves behind ionized sites. The presence of such dopant sites can act as traps for excited charge carriers\textcolor{black}{\cite{SiTauTempDep}}. The detrapping of charge carriers from such states limits the recombination process leading to significantly higher characteristic time for the recombination.

%\subsection{Sample thickness dependence of lifetime}

%EZ NEM JÖN EBBE A CIKKBE. EZT VEGYÜK KI

%For the characterization and demonstration of our measurement process we used a Silicon sample series made specifically for this purpose. We chose an n and a p type silicon wafer, cut out 10mmX10mm sized square shaped samples from them. We placed the samples into hydrofluoric acid to create samples with different thickness. This lead to similar sized samples with the same basic properties that differ in thickness. 

%We measured the recombination time of each sample with the help of the presented microwave photoconductance decay measurement setup and calculated the mobility of charge carriers in them. We also conducted a measurement of the recombination time with the tools of Semilab Semiconductor Physics Laboratory Co. Ltd.. The results are shown in Table \ref{thicknestable}. and show that the recombination times obtained from the measurements fall in line with the results made with the instrument of Semilab. We also calculated the mobility of charge carriers in the material from the measured recombination time and the diffusion coefficient that is known from the literature\cite{}. The mobility values we obtained fall in line with the literary values expected for n and p type silicon samples. This shows that the measurement system works correctly, the measured recombination times fall in line with the expectations.

%\subsection{Power dependence of the relaxation curve}

%\section{suplementary material}

\section{Conclusions}
\noindent
In conclusion, we presented a microwave photoconductance instrument with the capability of detecting time-resolved radiofrequency (or microwave) detected photoconductance decay and steady-state photoconductance. We use a phase-sensitive detection method utilizing an IQ mixer, which enables to study complex material parameters. We demonstrated the broadband measurement capabilities of the coplanar waveguide-based system and that it also enables temperature-dependent measurements of charge-carrier recombination time in the $20-300\ \text{K}$ range. When excited with two different laser wavelengths, the system allows to disentangle surface related and bulk recombination properties.

\section{Acknowledgements}
\noindent
The Authors are indebted to R.~Ga\'{a}l and T.~Pinel for the help with the instrumental setup. Work supported by the National Research, Development and Innovation Office of Hungary (NKFIH), and by the Ministry of Culture and Innovation Grants Nr. K137852, 2022-2.1.1-NL-2022-00004, 2019-2.1.7-ERA-NET-2021-00028 (V4-Japan Grant, BGapEng), TKP2021-NVA-02, C1010858 (KDP-2020) and C1530638 (KDP-2021).

%\bibliographystyle{IEEEtran}
%\bibliography{Bojtor_liter}
    

\end{document}